\documentclass[runningheads]{llncs}
\usepackage{graphicx}
\usepackage{mathtools}
\usepackage{xcolor}
\usepackage{soul}
\usepackage{color}
\usepackage{makecell}
\usepackage{amsmath,amsfonts}

\begin{document}
\title{Investigating Perceptions of Social Intelligence in Simulated Human-Chatbot Interactions}
\titlerunning{Social Intelligence in Simulated Human-Chatbot Interactions}
%
\author{Natascha Mariacher\inst{1} \and
Stephan Schl\"{o}gl\inst{1}\orcidID{0000-0001-7469-4381} \and
\\Alexander Monz\inst{2}}
\authorrunning{N. Mariacher et al.}
%
\institute{MCI Management Center Innsbruck, Innsbruck, Austria\\
Dept. Management, Communication \& IT \and 
Dept. Digital Business \& Software Engineering\\
\email{stephan.schloegl@mci.edu}\\
}
\maketitle              
%
\begin{abstract}
With the ongoing penetration of conversational user interfaces, a better understanding of social and emotional characteristic inherent to dialogue is required. Chatbots in particular face the challenge of conveying human-like behaviour while being restricted to one channel of interaction, i.e., text. The goal of the presented work is thus to investigate whether characteristics of social intelligence embedded in human-chatbot interactions are perceivable by human interlocutors and if yes, whether such influences the experienced interaction quality. Focusing on the social intelligence dimensions \textit{Authenticity}, \textit{Clarity} and \textit{Empathy}, we first used a questionnaire survey evaluating the level of perception in text utterances, and then conducted a Wizard of Oz study to investigate the effects of these utterances in a more interactive setting. Results show that people have great difficulties perceiving elements of social intelligence in text. While on the one hand they find anthropomorphic behaviour pleasant and positive for the naturalness of a dialogue, they may also perceive it as frightening and unsuitable when expressed by an artificial agent in the wrong way or at the wrong time.    

\keywords{Conversational User Interfaces \and Chatbots \and Social Intelligence \and Authenticity \and Clarity \and Empathy.}
\end{abstract}
%
%
%
\section{Introduction}
The ongoing success of mainstream virtual assistants such as Apple's \textsc{Siri}, Microsoft's \textsc{Cortana}, Google's \textsc{Assistant} or Amazon's \textsc{Alexa}, fosters the continuous integration of conversational user interfaces into our everyday lives. Beyond these speech-based intelligent agents, also text-based conversational interfaces have grown significantly in popularity~\cite{pinhanez2017design}. That is, in 2017 alone the number of chatbots offered on Facebook's \textsc{Messenger} platform has doubled\footnote{https://venturebeat.com/2017/04/18/facebook-messenger-hits-100000-bots/ [last accessed: March 11, 2019]}. To this end, Beerud Sheth, co-founder and CEO of Teamchat, states ~\textit{``we're at the early stages of a major emerging trend: the rise of messaging bots''}\footnote{https://techcrunch.com/2015/09/29/forget-apps-now-the-bots-take-over/?guccounter=2 [last accessed: March 11, 2019]}. So far, this trend is mainly visible in the customer support domain, where in recent years bot technology has gained significant ground (e.g.,~\cite{xu2017new,cui2017superagent,jacobs2017top}). In other commercial fields, however, it seems that the technology first needs to adopt more human-like traits for it to be accepted, particularly in areas such as shopping assistance, consulting or advice giving. That is, conversational agents may need to be recognized as social actors and, to some extent, be integrated into existing social hierarchies~\cite{wallis2005trouble}. Albrecht refers to this as the need for entities (both artificial and human) to be socially intelligent, which according to his \textbf{S}.\textbf{P}.\textbf{A}.\textbf{C}.\textbf{E} model encompasses the existence of a certain level of \textit{\textbf{S}ituational Awareness}, the feeling of \textit{\textbf{P}resence}, and the ability to act \textit{\textbf{A}uthentic}, \textit{\textbf{C}lear} as well as \textit{\textbf{E}mpathic}. Following his theory, the work presented in this paper aims to investigate whether (1) respective elements of social intelligence embedded into textual conversation are perceived by users; and (2) whether such perceptions may influence the experienced interaction quality. In particular, our work focuses on the dimensions \textit{Authenticity}, \textit{Clarity} and \textit{Empathy} embedded into human-chatbot interactions, leaving Albrecht's \textit{Situational Awareness} and \textit{Presence} for future investigations.
\section{Related Work}
In order to better understand elements of social intelligence and how they may be embedded into text-based human-chatbot interaction it seems necessary to first elaborate on the history and goals of intelligent agents as well as on previous work aimed at investigating traits of socially intelligent behaviour, and how such may be exhibited.

\subsection{Intelligent Agents}
Ever since its early days, research in Artificial Intelligence (AI) has worked on two branches, one focusing on using technology to increase rationality, effectiveness and efficiency~\cite{russell2016artificial}, and the other one concentrating on building machines that imitate human intelligence~\cite{brooks1991intelligenceRepresentation}. Speech recognition, autonomous planning and scheduling, game playing, spam fighting, robotics, and machine translation are just some of the accomplishments that can be attributed to the combined efforts of these two branches~\cite{russell2010socialIntelligence}. An achievement of particularly high impact lies in the field's steady progress made towards building autonomous machines, i.e., so-called agents which \textit{``can be viewed as [entities] perceiving [their] environment through sensors and acting upon that environment through actuators''} ~\cite[p. 34]{russell2010socialIntelligence}. In other words, those machines are able to perceive and interpret their surroundings and react to it without requiring human input~\cite{persson2001ias,wooldridge1994agents}.
A special form of autonomous systems may be found in conversational agents, which communicate or interact with human interlocutors through means of natural language~\cite{Shawar2003UsingDC}. Chatbots are a subcategory of these conversational agents defined, according to Abbattista and colleagues, as \textit{``[software systems] capable of engaging in conversation in written form with a user''}~\cite{abbattista2002improving}. They simulate human conversation using text (and sometimes voice)~\cite{deangeli2001unfriendlyUser,sha2009english}, whereupon their abilities are formed by an illusion of intelligence that is achieved through little more than rule-based algorithms matching input and output based on predefined patterns~\cite{deangeli2001unfriendlyUser,Shawar2003UsingDC}. In addition, these agents apply several tricks so as to steer and/or manipulate the conversation in a way which feigns intelligence~\cite{Mauldin1994CHATTERBOTSTA}. What they usually miss, however, are linguistic capabilities that show social rather than utilitarian intelligence.  
\subsection{From Intelligence to Social Intelligence}
Research focusing on intelligence has, for a long time, revolved around a single point of measurement, the \textit{Intelligence Quotient} (IQ). Recent developments, however, suggest that a multi-trait-concept represents reality more accurately~\cite{thorndike1920intelligenceUses,gardner1983,albrecht2006social}. Albrecht, for example, distinguishes between \textit{Abstract}, \textit{Social}, \textit{Practical}, \textit{Emotional}, \textit{Aesthetic} and \textit{Kinesthetic} intelligence. Concerning conversational agents, it is particularly \textit{Social Intelligence} (SI), described as \textit{``[...] the ability to get along well with others and to get them to cooperate with you.''}~\cite[p.3]{albrecht2006social}, which seems increasingly relevant.  
At its core, SI is understood as a determining factor that motivates people to either approach each other or distance themselves. It encompasses a variety of skills and concepts, such as an understanding of the behaviour of others ~\cite{baron-cohen1999fmri}, \textit{``role taking, person perception, social insight and interpersonal awareness''} ~\cite[p. 197]{ford1983}, adaption to the situational and social context ~\cite{albrecht2006social} and the engagement in \textit{`nourishing'} instead of \textit{`toxic'} behaviour. Based on these factors Albrecht proposes the S.P.A.C.E model which describes, as already highlighted, SI to be composed of \textit{Situational Awareness}, \textit{Presence}, \textit{Authenticity}, \textit{Clarity} and \textit{Empathy}~\cite{albrecht2006social}.

\subsubsection{Situational Awareness} -- The ability to understand people in different situations is described as \textit{Situational Awareness} (SA)~\cite{albrecht2006social}. It consists of knowledge about our environment and the subsequent extrapolation of  information that accurately predicts or initiates future events~\cite{so2004saias}. Any given situation is governed by several aspects, such as social rules, patterns and paradigms, the evaluation of behaviours~\cite{lei2005}, and the different roles within the relevant social hierarchies~\cite{schmidt2012achtsamkeit}. Therefore, it requires people to engage with others at an emotional level~\cite{albrecht2006social} and to correctly identify norms within social groups as well as emotional keywords~\cite{goleman2008leadership}.
Contextual information is herein crucial so as to correctly read and interpret a situation. This context can be described as \textit{``any information that can be used to characterize the situation of an entity''}~\cite[p. 5]{Dey2001} and further be divided into proxemic, behavioural, and semantic contexts. The proxemic context describes the physical space in which an interaction takes place, the behavioural context includes emotions and motivations of individuals during a given interaction, and the semantic context refers to language and associated meaning. Hence, when aiming to simulate human-human interactions, socially intelligent chatbots would need to consider these contextual attributes as well -- perceiving, comprehending and using them to project potential future statuses of the conversation~\cite{endsley1995situationalAwareness}.

\subsubsection{Presence} -- Albrecht's \textit{Presence} (P) dimension describes the way people use their physical appearance and body language to affect others. It is influenced by factors such as the first impression, a person's charisma, the respect they exhibit towards others, and the naturalness of their appearance~\cite{albrecht2006social}. In virtual environments or when communicating with artificial entities, however, \textit{Presence} is substituted by \textit{Social Presence} ~\cite{GEFEN2004,weisberg2011purchaseIntent}. The term is defined as \textit{``the degree of salience of the other person in the interaction and the consequent salience of the interpersonal relationships''}~\cite[p. 65]{short1976social}. In online settings, for example, presence can be created through measures such as the integration of people's names into conversations or by including pictures of smiling people on websites~\cite{GEFEN2004}. High social presence is particularly important in e-commerce, as it has shown to positively influence trust~\cite{GEFEN2004,weisberg2011purchaseIntent}. From a chatbot perspective it is especially the degree of anthropomorphism which influences how social presence is perceived~\cite{nowak2003Anthromorphism}. Yet, increasing levels of anthropomorphism do not always correlate with increasing levels of perceived presence, as overly realistic entities may quickly trigger suspicion or feelings of uneasiness often referred to as the uncanny valley effect~\cite{mori2012uncanny}.

\subsubsection{Authenticity} -- The concept of \textit{Authenticity} (A) encompasses the \textit{``notions of realness and trueness to origin''}~\cite[p. 457]{buendgens2014Authenticity}, uniqueness, and originality~\cite{gundlach2012authenticity}. Honesty and sincerity also indicate responsibility and empathy towards others~\cite{albrecht2006social}. Fulghum describes authenticity as the fundamental social rules of \textit{`playing fair'} and \textit{`sharing everything'}~\cite{fulghum1986kindergarten}. To this end, research has also emphasized the differences between authentic and rather simulated relationships~\cite{turkle2007authenticity}. In addition, particularly in  contexts where direct personal contact is lacking, trust becomes a vital component to authenticity, for it can easily be damaged and may consequently trigger doubts concerning a company's products and services~\cite{Chen2003ConsumerTrust}. In essence, the need for companies (and their brands) to act authentic has been steadily increasing ~\cite{Beverland2005aAuthenticity}, since \textit{`in times of increasing uncertainty, authenticity is an essential human aspiration, making it a key issue in contemporary marketing and a major factor for brand success.''}~\cite[p. 567]{Bruhn2012BrandAuthenticity}. Hence, a socially intelligent chatbot may need to also reflect a certain level of authenticity so as to not cause potentially harmful effects~\cite{Neururer2018_01}.

\subsubsection{Clarity} -- The ability to clearly express what people are thinking, what their opinions and ideas are, and what they want is represented by Albrecht's \textit{Clarity} (C) dimension. Expressing oneself clearly is not limited to the words and phrases used, but also encompasses how people speak, how they use their voice and how compellingly they express their ideas. Small differences in expression can evoke vast differences in meaning ~\cite{albrecht2006social}. Albrecht identified various strategies which are recommended to improve clarity, for example the \textit{``dropping one shoe''} strategy, in which one creates an expectation through a distinct message early in a conversation~\cite{albrecht2006social}. Other strategies include the use of metaphors or graphical explanations such as diagrams or pictures. As a general rule, one should be \textit{``mentally escorted''} so as to reduce cognitive load. Yet, pure text-based communication, as it is inherent to the use of chatbots, differs from face-to-face communication in several ways, most notably in style, spelling, vocabulary, the use of acronyms and emoticons~\cite{barton2013language}, and obviously the lack of additional information transmitted through mimical, gestical as well as voice related expressions. It seems therefore vital to take these factors into account when designing text-based interactions, and search for alternative ways of increasing clarity.

\subsubsection{Empathy} -- The final dimension of Albrecht's S.P.A.C.E. model is described by~\textit{Empathy}. The literature has no consensual definition of empathy so that various methods and scales of measurement exist~\cite{gerdes2010empathy}. A common understanding among this plethora of definitions is that emphatic behaviour means that people treat other people's feelings adequately~\cite{albrecht2006social} and consequently that they respond to them appropriately. Empathy further includes the skill of understanding a given situation and accepting the feelings of others, even though oneself may disagree with all or some of them~\cite{albrecht2006social,preece1999empathic}. As Decety and Moriguchi describe it:
\textit{``Empathy is a fundamental ability for social interaction and moral reasoning. It refers to an emotional response that is produced by the emotional state of another individual without losing sight of whose feelings belong to whom''}~\cite{decety2007empathic}.
Given that Empathy plays an important role in human-human interactions, we may argue that the transfer of empathic characteristics to human-chatbot interaction is of similar importance. Yet, modelling empathic behaviour requires assessing the context of social situations and consequently to determine which behaviour is required at what time~\cite{mcquiggan2007modeling} -- a rather challenging problem which is currently worked on by both academia and industry.

\vspace{0.3cm}

\noindent Although Albrecht's S.P.A.C.E model encompasses a total of five dimensions of SI, our initial investigations into human-chatbot interaction put their focus on only three of them; i.e., \textit{Authenticity}, \textit{Clarity}, and \textit{Empathy}. The other two, i.e., \textit{Social Awareness} and \textit{Presence}, may, however, be subject to future research.

\section{Methodology}
In order to investigate whether users (1) perceive elements of \textit{Authenticity}, \textit{C}larity and \textit{E}mpathy in chatbot interactions, and (2) whether such would effect their experience of the interaction, we followed a tripartite research methodology:

\subsection{Step 1: Evaluating SI Elements in Text}
As a first step, we used the literature on SI to design text utterances to be used by a chatbot in three different use cases scenarios (i.e., buying shoes, booking a flight, and buying groceries). For all three of these scenarios we created different sets of German text utterances. Set one focused on \textit{Authenticity}, set two focused on \textit{Clarity} and set three focused on \textit{Empathy} (note: there was some overlap between sets but the majority of utterances were distinct with respect to their inherent SI characteristics). The resulting 89 text utterances were then evaluated by N=55 students, all of whom were German native speakers, for their perceived degree of clarity, authenticity and empathy. That is, each of the utterances was rated for each of the three SI characteristics on a scale from \textit{1=low} to \textit{10=high}.\footnote{(note: the complete list of German utterances and their English translations, incl. information on elements of SI which was given to students, is available here: https://tinyurl.com/y4zjd5cz)}

\subsection{Step 2: Simulating Chatbot Interactions}
As a second step, we used a Wizard of Oz (WOZ) setup~\cite{dahlback1993wizard,SchloglS2015_02} to test our utterances in an interactive setting. To do so, we created three different Facebook profiles, all of which pretended to be chatbots yet were operated by a researcher using our pre-defined sets of text utterances. Profile one, called \textit{Johnny}, acted as the authentic chatbot (using the utterances designed to be authentic), \textit{Claire} acted as the one conveying clarity and \textit{Deni} as the empathic one. A total of N=18 German-speaking participants interacted with the simulated chatbots in the three afore mentioned interaction scenarios. Those scenarios were counterbalanced so that each participant interacted with each of the simulated chatbots following a different scenario, leaving us with six measurement points for each chatbot/scenario combination. The scenarios were as follows:

\paragraph{\textbf{Scenario 1: Buying Shoes} --}
\textit{You want to buy new shoes and decide to order them online. Your favorite webshop has developed a new Facebook \textsc{Messenger} chatbot to help you find the perfect model. Here is some information on what you are interested in:
\begin{itemize}
    \item Model: [Female: sneakers | Male: sport shoes]
    \item Color: Black
    \item Price: approx. EUR 90,-
    \item Size: [Female: 40 | Male: 46]
\end{itemize}
You already have an account with the webshop holding all relevant account data incl. payment information and delivery details. Note: Please complete the purchase only if you have found the perfect shoe for you. Otherwise, enjoy the interaction with the chatbot and ask anything you would like to know, even if it does not fit the scenario. Now, please click on the chatbot icon for [Johnny -- Claire -- Deni], and start the conversation with the sentence ``Start Conversation [Johnny -- Claire -- Deni]''. You can terminate the conversation at any time entering the phrase ``End Conversation''}.

\paragraph{\textbf{Scenario 2: Booking a Flight} --}
\textit{You are planning a trip to Hamburg. You decide to use the newly developed Facebook \textsc{Messenger} chatbot of your favorite travel portal to book a flight matching the following details:
\begin{itemize}
    \item Departure Airport: [X]
    \item Arrival Airport: Hamburg
    \item From: June 1\textsuperscript{st} 2018
    \item To: June 5\textsuperscript{th} 2018
    \item Price: approx. EUR 150,-
\end{itemize}
You already have an account with the travel portal holding all relevant account data incl. payment information and delivery details. Note: Please book the flight only if you believe it to be a good match. Otherwise, enjoy the interaction with the chatbot and ask anything you would like to know, even if it does not fit the scenario. Now, please click on the chatbot icon for [Johnny -- Claire -- Deni], and start the conversation with the sentence ``Start Conversation [Johnny -- Claire -- Deni]''. You can terminate the conversation at any time entering the phrase ``End Conversation''}.

\paragraph{\textbf{Scenario 3: Buying Groceries} --}
\textit{You invited your friend to eat spaghetti with tomato sauce at your place, but unfortunately you do not have time to go and buy the relevant ingredients. Your favourite supermarket offers an online order service which most recently was extended by a Facebook \textsc{Messenger} chatbot. You decide to use the chatbot to help you choose the best options for:
\begin{itemize}
    \item Pasta
	\item Tomato sauce
	\item Chocolate for desert (most favorite: [X])
\end{itemize}
You already have an account with the shop holding all relevant account data incl. payment information and delivery details. Note: Please only chose products that suit your needs. Otherwise, enjoy the interaction with the chatbot and ask anything you would like to know, even if it does not fit the scenario. Now, please click on the chatbot icon for [Johnny -- Claire -- Deni], and start the conversation with the sentence ``Start Conversation [Johnny -- Claire -- Deni]''. You can terminate the conversation at any time entering the phrase ``End Conversation''}.

\vspace{0.3cm}

\noindent After each of these scenarios (which ended either by fulfilling the simulated purchasing task or by stopping the conversation) participants were asked to complete a questionnaire assessing the chatbot's perceived level of \textit{Authenticity}, \textit{Clarity} and \textit{Empathy}.\footnote{(note: the post-scenario questionnaire and its English translation is available here: https://tinyurl.com/y4zjd5cz)} 

\subsection{Step 3: Exploring Interaction Experiences}
As a third and final step, all WOZ study participants were asked about their perceived experiences interacting with the three (simulated) chatbots. Interviews, which happened right after each WOZ experiment, were recorded, transcribed and analyzed employing Mayring's qualitative content analysis method~\cite{mayring2014qualitative}.\footnote{(note: the used interview guidelines and their English translations are available here: https://tinyurl.com/y4zjd5cz)}   

\section{Discussion of Results}
The goal of our analysis was to shed some light on the question whether chatbot users are capable of perceiving elements of social intelligence embedded in chatbot utterances. The initial questionnaire survey aimed at validating our utterance designs, whereas the WOZ study as well as the interview analysis focused more on the actual interaction with the chatbot and respective experiences. 

\subsection{Results Step 1: SI Elements in Text} \label{sec:questsurvey}
A total of N=55 students (26 female/29 male) were first given information on elements of SI and then asked to rate our 89 utterances according to their level of perceived \textit{Authenticity}, \textit{Clarity} and \textit{Empathy}. Results show that if asked on paper, and taken out of an actual interactive setting, people find it difficult to separate distinct elements of social intelligence in text. That is, our data shows high levels of clarity, authenticity and empathy with all the evaluated utterances. While \textit{Clair} (clarity) and \textit{Deni} (empathy) scored highest in their respective dimensions, \textit{Johnny} (authenticity) came in second, after \textit{Clair}. That is, \textit{Clair's} utterances were rated the clearest (M=7.56, Median=8, Mode=10) as well as the most authentic (M=7.34, Median=8, Mode=10) and \textit{Deni's} utterances were perceived the most empathic (M=7.07; Median=8, Mode=10). Yet, when looking at the highest ranked sentences in each category, \textit{Johnny} accounted for the top 10 sentences in authenticity whereas for clarity the top 10 were split between \textit{Clair} and \textit{Johnny}, hinting towards a strong connection between those two SI characteristics. 

A Pearson correlation confirmed a significant positive relation between \textit{Authenticity} and \textit{Clarity} (r=0.718, p\textless.001) as well as between \textit{Authenticity} and \textit{Empathy} (r=0.682, p\textless.001) and between \textit{Clarity} and \textit{Empathy} (r=0.629, p\textless.001). 


\subsection{Results Step 2: Simulated Interactions}
In order to investigate embedded elements of \textit{Authenticity}, \textit{Clarity} and \textit{Empathy} in chatbot interactions, we conducted a WOZ experiment~\cite{dahlback1993wizard}. As mentioned above, three different FB profiles served as simulated chatbot agents -- \textit{Johnny} as the authentic one, \textit{Claire} as the clear one and \textit{Deni} as the empathic one. A member of our research group acted as the wizard, controlling each profile and using its respective text utterances to interact with a total of N=18 participants (8 female/10 male; 20-30 years old).

\subsubsection{Turn-taking --}
Looking at the turn-taking behaviour, we found that for the first scenario (i.e., buying shoes), the average turn-taking was 25.22 (SD=7.73), for the second one (i.e., booking a flight) it was 28.28 (SD=8.43), and for the third one (i.e., buying groceries) it was 41.67 (SD=12.04). The significantly higher turn-taking rate in scenario three may be caused by the number of products participants had to buy. As can be seen in Table~\ref{tab:turntakingchatbots}, there were also turn-taking differences between the different chatbots. Yet, those differences were not strong enough to conclude that certain SI characteristics would increase or decrease turn-taking.

{\renewcommand{\arraystretch}{2}%
\begin{table}[h!]
\centering
\caption{Average turn-taking per scenario and chatbot.}
\begin{tabular}{l p{2.8cm} p{2.8cm} p{2.8cm}} 
\hline
Chatbot Profile & \makecell{Scenario 1 \\ \textit{Buying Shoes} \\ $\bar{x}$\textsubscript{turns}}  & \makecell{Scenario 2 \\ \textit{Booking a Flight} \\ $\bar{x}$\textsubscript{turns}} & \makecell{Scenario 3 \\ \textit{Buying Groceries} \\ $\bar{x}$\textsubscript{turns}} \\ 
\hline
\textit{Johnny} (authentic) & \makecell{28.00} & \makecell{27.17} & \makecell{45.17} \\ 
\textit{Claire} (clear) & \makecell{23.33} & \makecell{31.67} & \makecell{38.50} \\ 
\textit{Deni} (empathic) & \makecell{24.33} & \makecell{26.00} & \makecell{41.33} \\ 
\hline
\end{tabular}
\label{tab:turntakingchatbots}
\end{table}
}

\subsubsection{Post-scenario Questionnaire--}
Looking at the post-scenario questionnaire we asked participants to complete after the interaction with each of the simulated chatbots, it can be seen that \textit{Johnny} was the one most liked, which made him also the favorite when it comes to recommending a chatbot to a friend. \textit{Claire}, however, seemed to be the most helpful one. 
As for elements of SI, the questionnaire included three questions measuring perceived \textit{Authenticity} (i.e., Q5--7), three questions measuring perceived \textit{Clarity} (i.e., Q8--10) and three questions perceived \textit{Empathy} (i.e., Q11--13). Based on our intentions, \textit{Johnny} should have obtained the highest average scores regarding authenticity, yet results show that similar to the outcome of the questionnaire survey described in Section~\ref{sec:questsurvey} it was \textit{Claire} who was perceived the most authentic (M=6.24). Unfortunately, our intentions with respect to offering a distinctively clear and a distinctively empathetic chatbot were also not met, as both \textit{Clair} as well as \textit{Deni} received the lowest scores in their respective SI characteristics. 


\subsection{Results Step 3: Qualitative Feedback}
Finally, we asked participants about their perceptions of SI elements exhibited by the chatbots and in general about their interaction experiences. 

\subsubsection{Baseline Feedback --}
To gain some baseline insights with respect to participants' potential expectations, we first asked them about  elements of social intelligence found in traditional sales contexts (i.e., when interacting with human sales personnel in similar settings). To this end, an often named requirement was the ability to judge whether a customer actually requires help; e.g., P13: \textit{``[the person] should definitely notice if someone needs help -- when I know what I want, then I do not need help''}. Also, it was highlighted that sales personnel needs to show deep knowledge in their respective product category and thus should be able to provide relevant information. Further, rather obvious characteristics such as a certain level of politeness, respectfulness, openness, self-confidence and efficiency were named as elements which would accommodate a customer-friendly atmosphere and thus may add to the level of perceived social intelligence.

\subsubsection{General Chatbot Feedback --}
As opposed to their opinions regarding human-human interaction, our chatbot interactions seemed to polarize participants. While half of them did not see a reason for using a chatbot, the other half perceived the technology as a positive feature potentially helpful in a number of different e-commerce settings. In particular, as an alternative to the often offered search function. Surprisingly, however, participants did not perceive great differences between our chatbots. Overall, the majority found them friendly and polite, yet rather generic, static and potentially time consuming. Differences as to the use of words or sentence structures, although clearly existing, were hardly noticed. Consequently, when asked about their preferred scenario (or chatbot), participants supported their choice with scenario inherent characteristics, such as efficiency, information quality or topic interest, and did not refer to the linguistic behaviour of the chatbot. Hence, it was mainly the speed with which an interaction could be completed that let participants express their preference for one or another chatbot.

\subsubsection{Feedback Concerning Elements of Social Intelligence --}
Although participants did not perceive clear differences in chatbots' communication styles, they found elements of social intelligence, such as jokes or the use of methaphors, an important aspect adding to the naturalness of conversations. That is, the slightly anthropomorphic behaviour created a more pleasant and relaxed atmosphere in which a participant's counterpart was perceived to have its own personality. Also, if not overplayed, it increased a chatbot's trustworthiness; although, the tolerance between comfort creating and overuse of such human-like linguistic behaviour seems rather small, as herein participants' perceptions quickly changed from nice and funny to annoying, dumb and ridiculous.     

\subsubsection{Feedback Concerning Authenticity --}
Trying to understand what participants would expect from an authentic chatbot, we first asked them about their understanding of authenticity. Answers show that for them authentic behaviour means to follow goals despite predominant opposing opinions, to not `put on an act' for others, and to be honest and trustworthy. However, when asked about the characteristics of an authentic chatbot, participants were rather indecisive, so that the question whether a chatbot could be authentic or whether it would merely mimic human behaviour, remained unanswered. However, one characteristic that seemed to unite people's understanding of an authentic chatbot concerns its perceived honesty, both in that it is honest about its artificial nature and that it is honest with respect to the products it recommends. As this perception of honesty has to be earned, participants often suggested an increased level of transparency, particularly in the initial stages of an interaction. For example, showing a selection of products and their respective attributes, which would allow customers to see for themselves whether the allegedly cheapest product is really the cheapest. This aspect of allowing people to validate collected information also adds to a chatbot's trustworthiness -- a characteristic which, due to today's often predominant perception of technologies' increasingly manipulative nature, has to first be built up. Other mentioned characteristics which may add to the authenticity of a chatbot include its consistent use of words and language structures, its ability to answer non-contextual questions, or it addressing other interlocutors with their name.   

\subsubsection{Feedback Concerning Clarity --}
Clarity was defined as the ability to express oneself clearly and in an understandable manner. Such was also transferred to chatbot behaviour. For example, participants found it important that chatbot answers are coherent and not too long, yet at the same time do not look too basic or scripted. Also, in terms of clarity a chatbot should refrain from asking several pieces of information at once, as our participants were often unsure about how to answer such a request, trying to compose a sentence that holds many pieces of information while still being simple to comprehend. A lack of understanding as to how chatbot technology works and consequently `makes sense' of information generates the perception that inquiries as well as answers have to be simply structured. Consequently, a request for multiple pieces of information created an uneasiness in people as they felt unsure about the main purpose of the inquiry.

\subsubsection{Feedback Concerning Empathy --}
To be empathic means to be capable of understanding how other people think and feel. To be able to recognize the emotions of other people and to respond appropriately to the observed behaviour. It also incorporates the acceptance of other viewpoints and the showing of interest in other people and their particular situations. Given this need for understanding and recognition of behaviour and/or feelings, participants found that chatbots are not yet sufficiently developed to act emphatic. That is, people usually do not express themselves in text so clearly and extensively that a chatbot (or even another person for that matter) would be able to read their emotions. Furthermore, text-based chats do not really support the use of spontaneous gestures (besides the use of gesture-like elements found in different emoticons), yet gestures are an essential part of expressing emotions and consequently empathic behaviour. Even if certain language elements such as jokes or motivational phrases may convey feelings in conversational behaviour, and potentially trigger positive emotions, they were not perceived as being empathic (although famous early work has shown that people may attribute empathy to a chatbot therapist~\cite{weizenbaum1966eliza}). It was further stressed that there is a link between empathy and authenticity, for a chatbot that is not perceived  authentic can not be perceived empathic, as its alleged empathy could not be taken serious. A rather simple aspect of empathic chatbot behaviour, however, was found in reacting to people's concrete needs; e.g., in an e-commerce setting if a product is out of stock to offer an alternative product that is available. Also, showing interest i.e., continuous questioning, may add to a certain level of perceived empathy. Yet, also here the right level of interest is important, as otherwise human interlocutors may feel skepticism, wondering about the technology's intentions. Finally, utterances which seemingly create positive feelings, such as positive surprises, compliments or jokes, do also add to an overall empathic impression. Although, it has to be mentioned that to this end individual perceptions can be heavily subjected to one's current mood and emotional situation. While in human-human interaction an interlocutor may be able to react and consequently adapt his/her behaviour to an interlocutor's mood, current chatbots are usually not equipped with such an ability.  

\section{Reflection}
The study results presented above paint a rather divers and inconclusive picture of the current situation. That is, it seems that people  are able to relate to characteristics of SI when interacting with humans, yet they find it difficult to transfer this understanding to chatbots. In spite of our participants' predominantly positive ratings with respect to the conversations and their strong believe in the potential the technology holds, they exhibited rather critical attitudes towards chatbots. We consistently found that they were rather goal-driven in their interactions, for which efficiency was perceived more relevant than socially intelligent interaction behaviour. While such may be owed to the chosen scenarios (i.e., situated in the online shopping domain), participants seemed to generally have low expectations in chatbot behaviour and thus did not believe the technology would be capable of handling and expressing human traits. Nonetheless, when asked about their concrete perceptions in more detail, impressions of socially intelligent chatbot behaviour were apparent. With respect to the perception of and consequent reaction to traits of SI in human-chatbot interaction we may thus conclude:      

\paragraph{\textbf{SI Utterances in General} --} The additional utterances we used to express elements of SI, were perceived positive, so that according to participants conversations were `nicer' and more `personal'. Mostly, they improved the linguistic exchange by making the interaction clearer, slightly more empathic (if one can speak of empathy when interacting with an artificial entity) and thus also more authentic. However, when used incorrectly, with respect to timing and/or contextual fit, they were quickly perceived as annoying and ridiculous -- probably more so than when knowingly interacting with a human being. Consequently one could argue that on the one hand our participants did not believe in complex human traits exhibited by chatbot technology, on the other hand, when exposed to this behaviour they were more critical concerning mistakes than they would have been with a human interlocutor.

\paragraph{\textbf{Authenticity} --} Although participants could not define what it would mean for a chatbot to be authentic, they found the interaction more pleasant when the chatbot showed anthropomorphic behaviour. So it does seem that they appreciated the perception of talking to a human being, although they knew that it was a chatbot with which the were conversing. To this end, it was particularly the impression of the system being honest and transparent which helped convey authentic behaviour. Also, it helped increase trust, which may be seen as a key element influencing a technology's success~\cite{mcknight2011trust}. 

\paragraph{\textbf{Clarity} --} In terms of clarity, participants were satisfied with the way the chatbots expressed themselves. For them the conversations were clear and understandable, and the used language was pleasant. However, participants interpreted clarity as an element of efficiency; i.e., whether the chatbot conversed in a way so that a swift goal completion was supported. Clarity in the sense of providing a better, more comprehensive understanding of the context and thus an increased level of information, seemed less relevant.  

\paragraph{\textbf{Empathy} --} Empathy was the one dimension of SI which was most criticized. In general, participants had difficulties believing that chatbots (or technology in general) may be capable of showing empathic behaviour, as relevant emotional input and its accurate interpretation are still missing. As for the emotional output, chatbot conversations were perceived as too short and transient to exhibit empathy. Those perceptions may also depend on a person's current emotional state and whether somebody is in the mood of conversing with an artificial entity (and consequently may engage in a more interactive dialogue that offers room for increasingly reflective behaviour). A final aspect of empathy may relate to its creation of affinity and subsequent emotional binding. Here previous work has shown that people build close connections with technologies such as mobile phones~\cite{vincent2005emotional}, robots~\cite{ogawa2008itaco} or even televisions  appliances~\cite{reeves1996people}, which go as far as to fulfill some sort of companionship role~\cite{benyon2010human}. Since a recent study showed that people ended up personifying Amazon's \textsc{Alexa} after only four days of using it~\cite{lopatovska2018personification} it may seem plausible that similar effects are inherent to human-chatbot interaction, for which empathy may in fact play a greater role than currently perceived by our study participants. 

\vspace{0.3cm}

\noindent Concluding, we may thus argue that our analysis of the perceptions and consequent evaluations of SI in human-chatbot interactions produced two important findings. First, we have seen that people are (for now) mainly interested in efficiency aspects when conversing with an artificial entity. That is, they primarily want to reach their goal. Seen from this perspective, the SI of chatbots seems to be of rather secondary importance. On the other hand, however, we have also seen that people often take elements of SI for granted. That is, they assume that chatbots understand how to use these elements to make a conversation more pleasant and natural. Consequently we might conclude that although people are often skeptical about human traits expressed by technological entities, they value a certain level of human-like behaviour.   
 
\section{Summary and Outlook}
Considering the increasing uptake of chatbots and other AI-driven conversational user interfaces it seems relevant to not only focus on the utilitarian factors of language but also on its social and emotional aspects. The goal of the here presented work was therefore to investigate whether traits of social intelligence implemented by human-chatbot interactions were perceivable by human interlocutors, and if yes whether such affects their experience of the interaction. Explorations employed a questionnaire survey (N=55) aimed at evaluating people's perceptions of various SI characteristics (i.e., \textit{Authenticity}, \textit{Clarity} and \textit{Empathy}) integrated into simple text utterances, and N=18 WOZ sessions incl. subsequent interviews, which aimed a investigating those characteristics and their perceptions in a more experimental setting. 

Results show that people have great expectations with respect to the conversational behaviour of chatbots, yet characteristics of SI are not necessarily among them. While the feeling of talking to a human being (i.e., anthropomorphic behaviour) is perceived as pleasant and may help increase transparency and consequently trust, people do not believe that a chatbot is capable of conveying empathy. Here it is mainly the reflective, non-goal oriented part of empathy which seems out of scope. While a chatbot may very well understand and react to a person's needs and desires in a given context, the interaction is usually to ephemeral to build up any form of deeper connection. Also, while empathic behaviour embedded in single text utterances is usually well received if appropriate, it is highly criticized when out of place - even more so than with human interlocutors.    

A limitation of these results can certainly be found in the exploratory nature of the setting. On the one hand our questionnaire survey evaluated the perception of text utterances that were out of context, on the other hand our WOZ-driven interactions focused on common online shopping scenarios, for which socially intelligent interaction behaviour may not be needed or even appropriate (note: online shopping also happens without conversational user interfaces). Thus, future studies should focus on a less structured situation, potentially building on the early work of Weizenbaum's Eliza~\cite{weizenbaum1966eliza}. Another aspect which was not subject to our investigations, yet may influence the perception of socially intelligent chatbot behaviour, concerns the adequate use of emoticons. As previous work has shown that emoticons trigger concrete emotions~\cite{yuasa2006emoticons}, we believe that a better understanding of when an how to use illustrative language in human-chatbot interaction is needed if we aim to eventually increase the level of social intelligence conveyed by these artificial interlocutors.

\section{Acknowledgments}
The research presented in this paper has been supported by the project EMPATHIC that has received funding from the European Union's Horizon 2020 research and innovation programme under grant agreement No 769872.

\bibliographystyle{splncs04}
\bibliography{wirn}
\end{document}